\documentclass{elsart}

\usepackage{graphicx}
\usepackage{amssymb,latexsym,times}

\begin{document}
\newcommand {\sax} {{\it Beppo}SAX }
\newcommand {\rosat} {{ROSAT }}
\newcommand {\rchisq} {$\chi_{\nu} ^{2}$} 
\newcommand {\chisq} {$\chi^{2}$}
\newcommand {\als} {$\alpha_1$}
\newcommand {\alh} {$\alpha_2$}
\newcommand {\ergs}[1]{$\times10^{#1}$ ergs cm$^{-2}$ s$^{-1}$}
\newcommand {\E}[1]{$\;\, \times10^{#1}$}
\newcommand {\gpeak} {$\gamma_{peak}$ }
\newcommand {\nupeak} {\nu_{peak} }
\newcommand {\es} {1ES 1426+428 }
\newcommand {\ax}{\alpha_x}
\newcommand {\C}{\^Cerenkov }
\newcommand {\ergcms} {erg cm$^{-2}$ s$^{-1}~$}

\begin{frontmatter}


\title{The SED of the TeV BLLac 1ES 1426+428
 after correction for the TeV--IR absorption}

\author[mpik]{L. Costamante}
\author[mpik]{F. Aharonian}
\author[merate]{G. Ghisellini}
\author[mpik]{ D. Horns}

\address[mpik]{Max Planck Institut f\"ur Kernphysik,  
Saupfercheckweg 1, D-69029  Heidelberg, Germany}
\address[merate]{Osservatorio Astronomico di Brera, Via Bianchi 42, 
I-23807 Merate, Italy} 
\vspace*{-0.5cm}
\begin{abstract}
The recent HEGRA detection and spectrum of 1ES 1426+428 at TeV energies,
once corrected for absorption using  present estimates of the diffuse 
extragalactic
IR background, suggest that the high energy peak of the Spectral Energy Distribution
(SED) could be much higher than the synchrotron one ($L_c/L_s>10$), and 
lie at energies above 8-10 TeV.
To see if such an SED could be accounted for, we have applied a
"finite injection time" SSC model, 
and present here some preliminary results.
Within this model, we found the need of an external ("ambient") contribution
to the energy density of seed photons, in order to account for both
the high Compton dominance and the hard spectrum.
\end{abstract}
\end{frontmatter}

\vspace*{-1.cm}

\section{Introduction}
\vspace*{-0.7cm}

The very high energy emission from extragalactic objects 
(and BL Lacs in particular) is likely affected by absorption 
due to the $\gamma$-$\gamma$ collision and pair production process
on the diffuse extragalactic IR background radiation (IRB).
%
%
Since the expected optical depths are energy--dependent and 
can be very high (see Fig. 1),
the original spectra are expected to be modified both in shape and intensity.
Although this gives the opportunity for an  independent 
measurement of the IR background on cosmological distances, 
it also constitutes a problem, since  
at present we cannot disentangle the two contributions: 
to constrain the IRB we should model the initial incident spectrum, but to model
the original blazar spectrum we need to know the amount of absorption, since also
in the most simple SSC models the physical parameters derived from the synchrotron 
peak are not sufficient to univocally constrain the full blazar SED, 
even not considering variability effects.
Simultaneous X-ray -- TeV observations and data at energies not (or less) 
affected by absorption (e.g. $<$ 100$-$200 GeV) are needed, as well as more 
TeV sources located at different redshifts. \\
%
Even if the uncertainties on the IRB intensity and shape are still very large
\cite{haus},
it is nevertheless interesting to see how the intrinsic blazar
spectra would be according to  present estimates of the IRB,
and if our emission models can account for them.
In this respect, the case of the extreme BL Lac 
1ES 1426$+$428 is of particular interest, since it is the TeV source with the highest redshift known
(z$=$0.129), and its spectrum has been measured by HEGRA up to 10 TeV 
\cite{ah,horns},
i.e. at energies where the optical depth is expected to be very high (see Fig 1, right). \\
As a working hypothesis, therefore, we have adopted two possible models for the IRB
(Fig.1 left) to reconstruct the 1ES 1426$+$428 intrinsic TeV spectrum,
and studied the resulting global SED 
applying a ``finite injection time" SSC model, which is  fully described in \cite{gg}.

\begin{figure}
\vspace*{-0.5cm}
\begin{center}
\resizebox{\hsize}{!}{\includegraphics*[width=9cm]{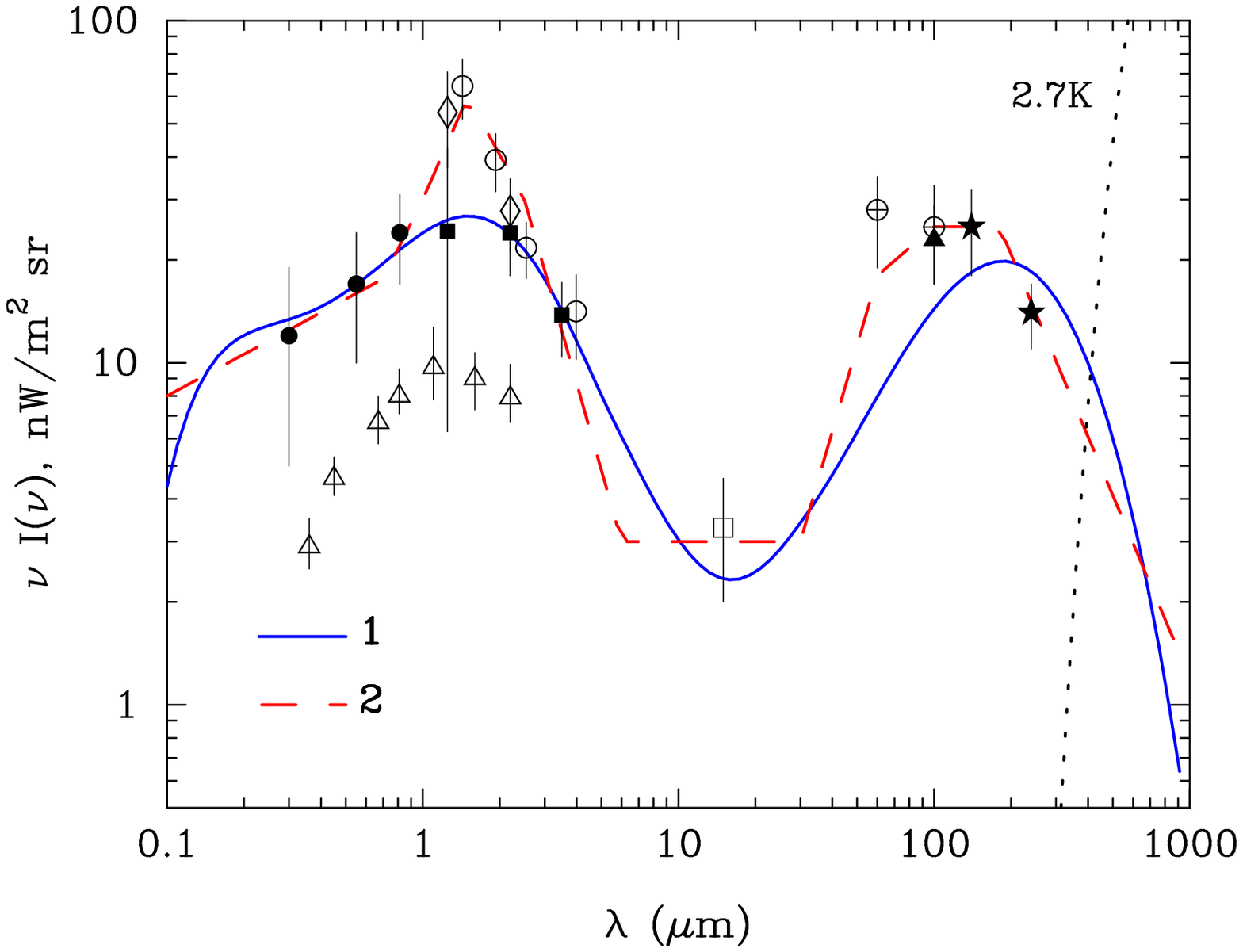}
 \includegraphics*[width=7cm]{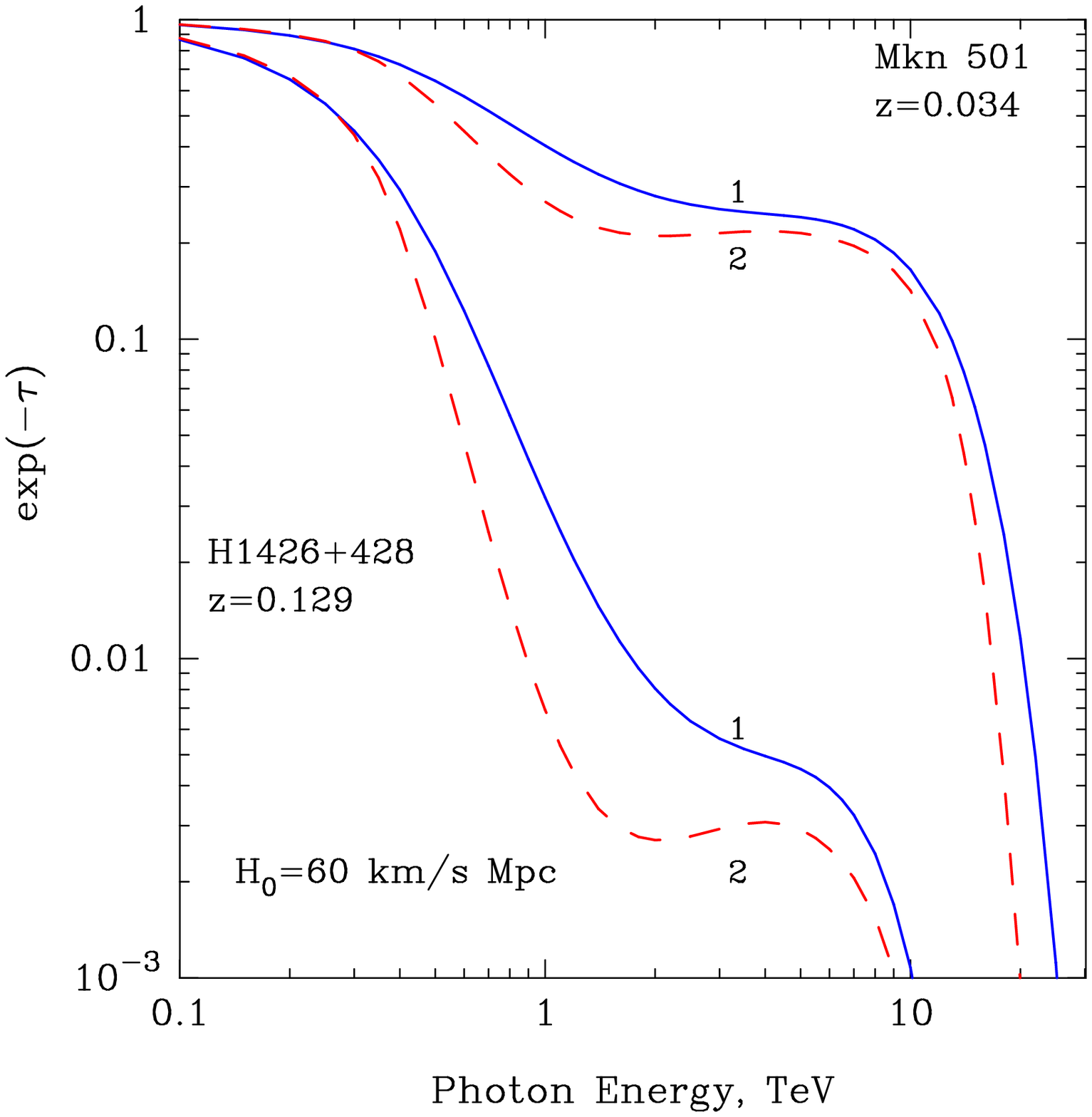} }
\vspace*{-0.8cm}
\caption{\footnotesize 
Left: SED of the IRB, according to published data, along with two
models in agreement with them: model 1 is based on calculations by 
Primack \cite{primack}, 
model 2 is designed to match the high fluxes 
below 2 $\mu$m (details in \cite{ah}). Right: 
The corresponding optical depths (expressed as exp($-\tau$)) 
for 1ES 1426+428 and Mkn 501. }
\end{center}
\end{figure}

\vspace*{-1cm}
\section{The TeV spectrum of \es }
\vspace*{-0.7cm}  
\es is a low power, high synchrotron peak ($\nupeak>100$ keV) BL Lac,
according to the most recent \sax observation 
(Feb. 1999, $\ax =0.9$, \cite{co}).
Fig. 2 shows its SED in the TeV range, with data from HEGRA \cite{ah}
(1999-2000, 1-10 TeV, re-analyzed for better 
energy resolution and sensitivity at high energies \cite{horns}), 
Whipple \cite{hor,petry} (2001, 0.3-2 TeV) and CAT \cite{dj} (1999-2000, 0.3-1 TeV). \\
Fig. 2 shows also the corresponding absorption--corrected points 
(using IRB ``model 1")
and power-law model spectra before and after absorption. \\
With both IRB models, the large optical depths yield intrinsic
fluxes  $>10^{-10}$\ergcms with hard spectra ($\alpha<1$), 
implying a Compton peak above 10 TeV with an apparent
luminosity $>10^{46}$\ergcms.    
If we assume an X-ray flux during the TeV observations 
not far from the historical data (as suggested by the XTE ASM lightcurves),
the resulting global SED (Fig. 3) seems characterized  
by a high Compton dominance ($L_c/L_s>10$) and a relatively high apparent 
bolometric luminosity ($>10^{46}$\ergcms): these results are not expected in 
the standard ``blazar sequence" scenario 
\cite{fo},
for which objects with high peak frequencies should have low bolometric 
luminosities ($<10^{45}$\ergcms) and a $L_c/L_s\sim1$. \\
%
%
%
%
\begin{figure}
\vspace*{-0.8cm}
\begin{center}
\resizebox{15cm}{!}{\includegraphics*[angle=-90,width=10cm]{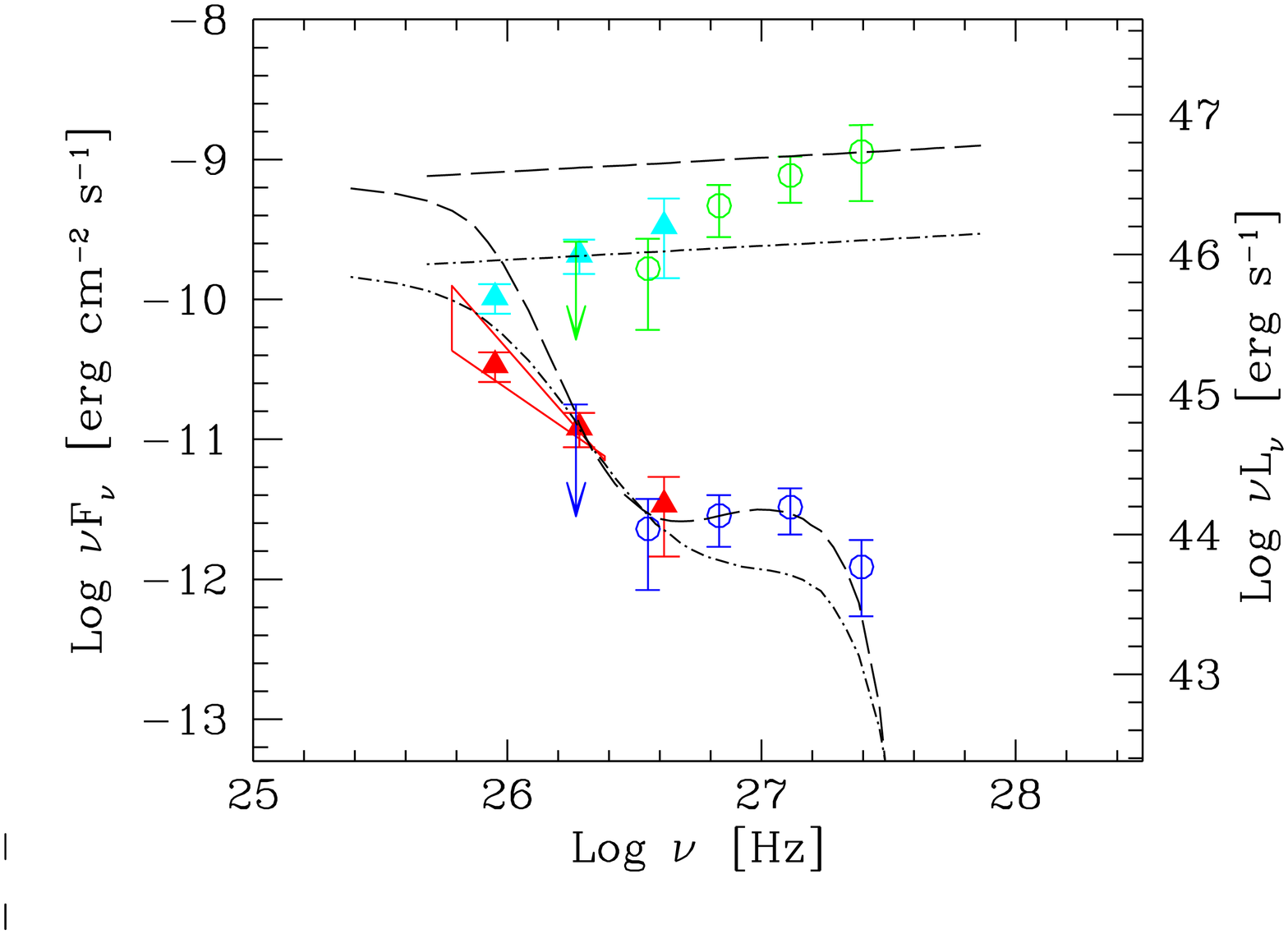}
 \includegraphics*[angle=-90,width=10cm]{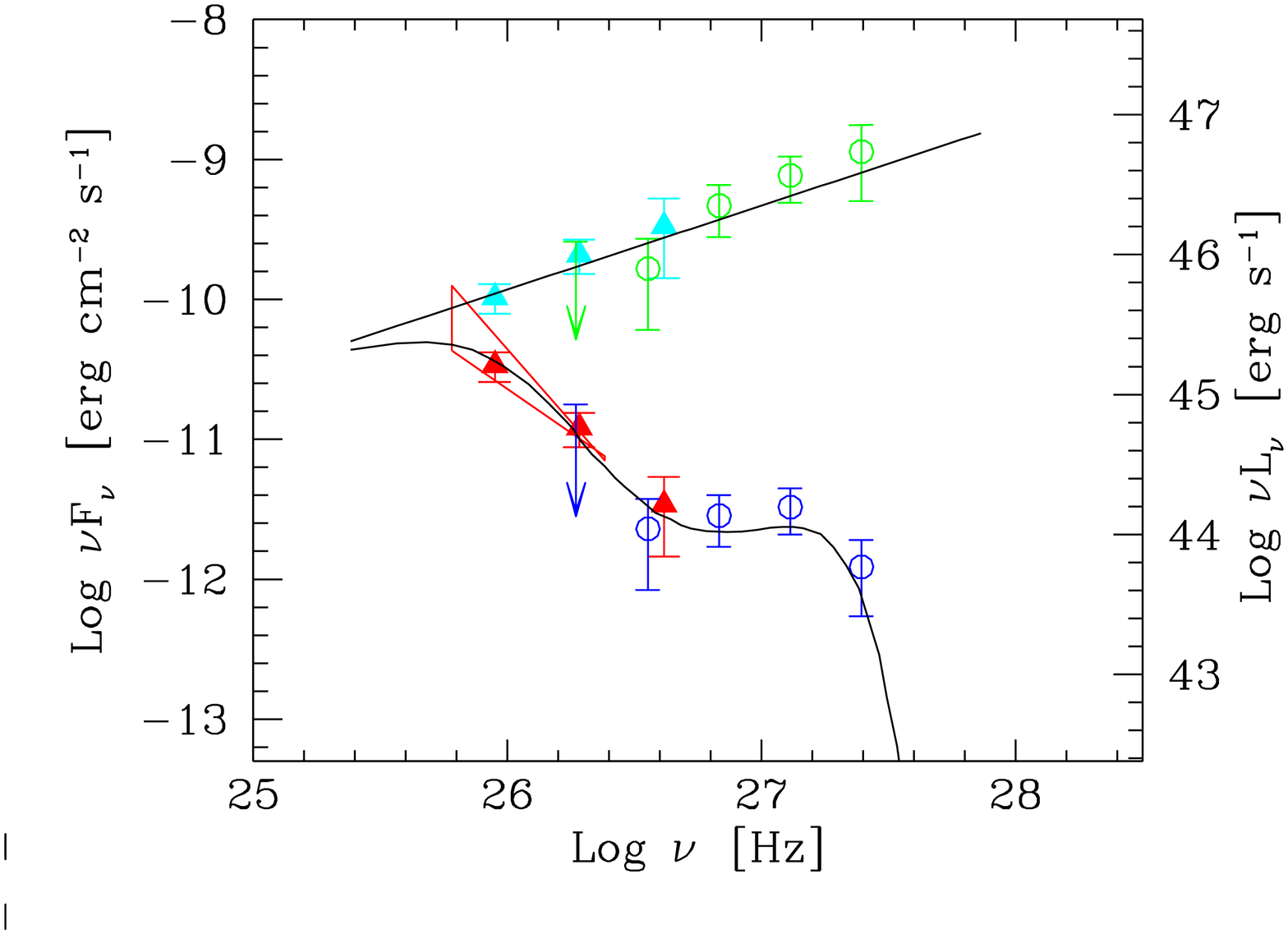} }
\vspace*{-0.8cm}
\caption{\footnotesize
The SED of 1ES 1426+428 in the TeV range. 
Open circles: HEGRA data. Triangles: Whipple data.
Bowtie: CAT data. The upper points are those
corrected for absorption using IRB ``model 1" 
(which gives less absorption than model 2, see Fig. 1).
The indication is for hard ($\alpha<1$) intrinsic spectra.
Left: model spectra  obtained absorbing a $\alpha=0.9$ power-law 
(i.e. the \sax slope)  with IRB model 1 (point-dashed line) and 2 (dashed line). 
Such spectra seem not able to account simultaneously for both HEGRA and Whipple data (\rchisq$>2$).
A possible solution is provided by harder intrinsic spectra (right, $\alpha=0.4$,           
``model 1" absorption, \rchisq$=0.8/7$ d.o.f.). Alternatively, a different shape
of the IRB is required, in particular between 2-10 $\mu$m 
(which affects the multi-TeV HEGRA data) and/or below 1$\mu$m 
(which affects the Whipple and CAT data).}  
\end{center}
\vspace*{0.1cm}
\end{figure}
To see if such an SED could be accounted for, we have tried
various fits with our SSC model, both assuming or not the \sax X-ray slope
also in the TeV range.
%
%
Within this model,
the pure SSC fits (Fig. 3 left) seem not able to provide at the same time
high Compton luminosities and hard spectra: 
the decline of the cross section in the Klein-Nishina regime
prevents the TeV electrons from scattering efficiently the synchrotron peak photons,
while the scattering in the more efficient Thomson regime occurs with photons
of lower energy density (optical and IR), yielding steep spectra in the TeV range.
These effects can be counterbalanced (Fig. 3 center and right) 
assuming an additional contribution to the energy density of the seed photons 
in the Thomson regime,
which we have preliminarly parameterized as a black body spectrum
peaked at $\nu_0\approx 2\times10^{13}$ Hz (in the comoving frame).
This ``ambient" contribution could be supposed to originate from
other parts of the jet (e.g. in a  spine-layer scenario \cite{ca})
or outside the jet itself. Model parameters are reported in Table 1. 

Concluding, if the assumed IR background will be confirmed,
the intrinsic SED of \es will require a  revision of 
the ``standard" blazar sequence  scenario. However, our SSC model
seems still able to account for such hard and luminous TeV spectra, 
but with the addition of an external contribution to the seed photon energy density.  
In the near future, we hope to resolve 
the ambiguity in the interpretation of TeV data with the new generations
of \C Telescopes (e.g. H.E.S.S.): their lower energy thresholds ($<0.1$ TeV) 
will provide us with data from both the absorbed and (almost) not absorbed part 
of the spectrum, and their better sensitivity will permit more detailed
studies of the correlated X-ray -- TeV variability,
allowing to more reliably model the full BL Lac SED and thus to constrain the IRB.

\begin{figure}
\vspace*{-2cm}
\begin{center}
\hspace*{-1cm}
\includegraphics*[angle=-90,width=17cm]{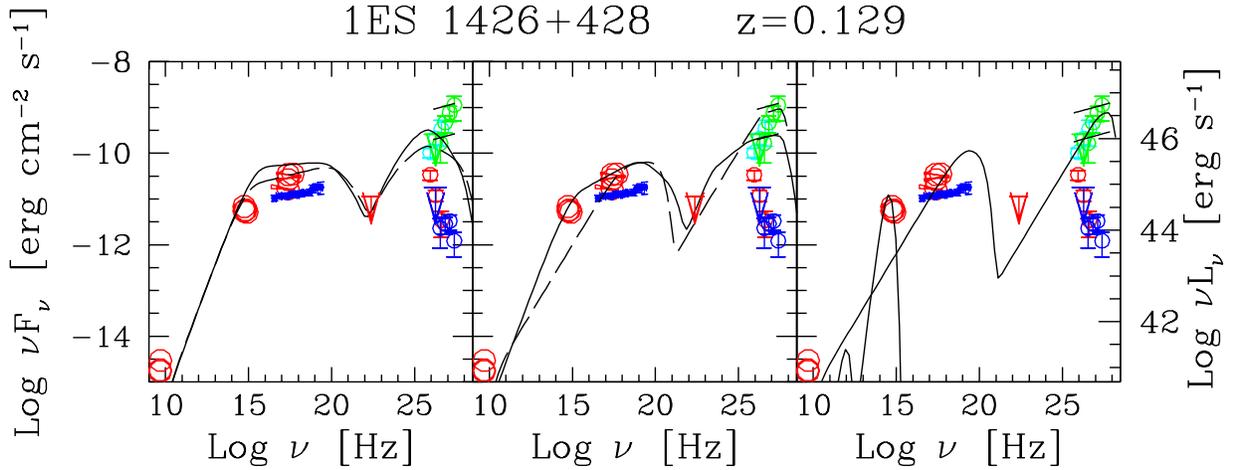}
\vspace*{-4.5cm}
\caption{\footnotesize  
The full SED of 1ES 1426+428.  The $\gamma-$ray data are those of Fig. 2; 
the two segments correspond to the two intrinsic powerlaws of Fig. 2
(left), and have the same slope of the \sax data. 
Left: pure SSC fits. Center: fits with an added external contribution of 
``ambient" seed photons (see text). Model parameters in the table below.
Full lines correspond to the "ssc1" and "amb1" fits, dashed lines
to the "ssc2" and "amb2" ones.
Right: another ``ambient" fit (``amb3"), but aimed
to reproduce the ``hardest spectrum" hypothesis of Fig. 2 (right), 
thus not considering the \sax slope.  In this case the
external contribution is also plotted, as it would appear if originating
inside the jet from a region with bulk motion $\Gamma_{amb}=\Gamma/2$; 
or outside the jet, from a spherical region (like the Broad Line region) 
of radius $R_{amb}\sim4\times10^{17}$ cm.
}  
\end{center}
\end{figure}

\begin{table}[t]
\begin{center}
\vspace*{0.2cm}
\footnotesize
\hspace*{-0.5cm}
\begin{tabular}{|l|cllllllllc|}
\hline 
 models &$L^\prime$ (erg s$^{-1}$)   &$R$ (cm)  &$B$ (G) &$\Gamma$ &$\theta$ &$s$ &$\gamma_{min}$ &
$\gamma_{\rm peak}$ &$\gamma_{max}$ & $U_{ext}/U_{sync}$  \\
%
%
%
\hline
ssc1   & 1.0e42  &  2.5e16  & 0.053  & 17  &  2 & 2.94  & 3.0e4 & 4.0e6 & 3.0e7 &  0 \\
ssc2   & 1.0e42  &  2.5e16  & 0.045  & 17  &  2 & 2.82  & 2.0e4 & 5.8e6 & 4.0e7 &  0 \\ 
amb1   & 3.0e42  &  2.5e16  & 0.045  & 17  &  2 & 2.60  & 2.0e4 & 1.9e6 & 2.0e7 & 0.13 \\
amb2   & 3.0e42  &  2.5e16  & 0.040  & 17  &  2 & 1.80  & 2.0e2 & 8.0e6 & 8.0e6 &  0.71 \\
amb3   & 3.0e42  &  3.4e16  & 0.030  & 19  &  2 & 1.80  & 2.0e2 & 7.0e6 & 7.0e6 & 0.29 \\
\hline 
\end{tabular}
\end{center}
\vspace*{0.2cm}
{\footnotesize Table 1. Parameters for the Fig. 4  fits,
namely: the intrinsic source luminosity, size of the emitting region, magnetic 
field, bulk Lorentz factor, viewing angle, injected slope of the particle distribution,  
min and max Lorentz factors of the injected electrons, Lorentz factors of the 
electrons emitting at the peak and 
the ratio of the "ambient" to the locally produced synchrotron energy density, 
in the comoving frame. }
\vspace*{0.2cm}
\end{table}

\end{document}